\documentclass[]{spie}  %>>> use for US letter paper
\usepackage{amsmath,amsfonts,amssymb}
\usepackage{graphicx}
\usepackage[colorlinks=true, allcolors=blue]{hyperref}

\begin{document}

\title{Implementation of nonlocal multi-photon interference\\ by mode swapping}

\author[a]{Holger F. Hofmann}
\author[a]{Yuki Kodama}
\author[b]{Jonte R. Hance}
\affil[a]{Graduate School of Advanced Science and Engineering, Hiroshima University,
Kagamiyama~1-3-1, Higashi Hiroshima 739-8530, Japan}
\affil[b]{School of Computing, Newcastle University, 1 Science Square, Newcastle upon Tyne, NE4~5TG, UK}

\authorinfo{Holger F. Hofmann: E-mail: hofmann@hiroshima-u.ac.jp}

% Option to view page numbers
\pagestyle{empty} % change to \pagestyle{plain} for page numbers   
\setcounter{page}{301} % Set start page numbering at e.g. 301

%%\begin{document} 
\maketitle

\begin{abstract}
Multi-photon interference can be observed using independently generated photons as input. In the most simple case, these photons meet up at a beam splitter, resulting in quantum interference between transmission and reflection of the photons. Here, we show that non-local multi-photon interference can be implemented by using a mode swap operation to generate entanglement between the photons detected in the outputs of two spatially separated interferometers. The spatial separation of the output photons makes this implementation of multi-photon interference particularly suitable for quantum protocols that distribute quantum information to different parties. 
\end{abstract}


\section{INTRODUCTION}
\label{sec:intro}  % \label{} allows reference to this section

The vast majority of optical quantum information technologies relies on multi-photon interferences to generate and manipulate the non-classical properties of light \cite{Pan12}. The textbook example for such multi-photon interference effects is the destructive interferences between two-photon transmission and two-photon reflection in the well-known Hong-Ou-Mandel effect \cite{Hon87}. This effect has been integrated into a wide variety of optical quantum circuits. Of particular interest is the generation of entanglement in post-selected circuits, either by heralding \cite{Hof01,San06,Oka09} or by output post-selection \cite{Ral02,Hof02,Bri03,Lan05,Kie05,Oka05}. When such operations are applied to larger numbers of photons and modes, they are usually analyzed in terms of multi-partite entanglement, where photon polarization is used to define a qubit and each spatial output path is considered to be a separate physical system \cite{Ono17,Kum23}. However, it is also possible to define optical systems as local multi-mode systems, where one set of modes is sent to location A and another set to a separate location B. In this manner, optical bipartite entanglement can be scaled up to arbitrary numbers of photons and modes \cite{Wu17,Kiy20}. If the latter definition of optical systems by the physical location of the optical modes is considered, the exchange of a specific set of modes between the two systems corresponds to a nonlocal interaction between the multi-mode systems involved in the exchange. Such a mode swap operation then constitutes an entangling interaction that can implement a nonlocal version of multi-photon interference. In the following, we will introduce a basic optical circuit for a nonlocal implementation of the Hong-Ou-Mandel effect, where photon bunching correlates the output mode of a photon in A with the output mode of a photon in B. This effect happens even though post-selection guarantees that the two photons never occupy the same mode or enter the same beam splitter. Mode swapping thus demonstrates that multi-photon interference is an effect caused by the non-local phase coherence between spatially separated modes.

\begin{figure}[ht]
%%\vspace{-1cm}
\begin{picture}(500,240)
%%\put(0,0){\framebox(500,240){}}
\put(20,0){\makebox(400,200){\vspace{-2cm}
\scalebox{0.8}[0.8]{
\includegraphics{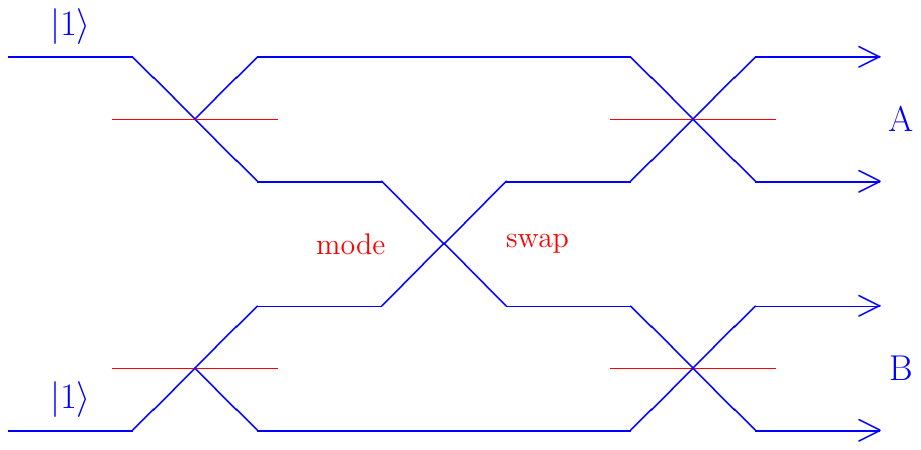}}}}
\end{picture}
%%\vspace{-5cm}
\caption{\label{setup}
Schematic representation of a basic mode swapping circuit. The upper two paths form the local system A and the lower two paths form the local system B. If one photon is observed in each system, the two photons have either stayed on their side or crossed paths without interacting.
}
\end{figure}

\section{Entanglement generation by mode swapping}

Optical qubits can be realized by a single photon in two modes. Here, we consider only spatial modes that can be represented by the optical paths of an interferometric setup. Qubit A is defined by one photon in two modes at location A, and qubit B is defined by one photon in two modes at location B. A mode swap operation interacting the two qubits is realized by exchanging one of the two modes in A with one of the two modes in B. A schematic representation of this setup is shown in Fig. \ref{setup}. Note that the two modes are swapped without any physical interaction. Interference effects are obtained by inserting the mode swap into a pair of local Mach-Zehnder interferometers, so that the state before the swap operation is
\begin{equation}
\mid \psi_1 \rangle = \frac{1}{2}\left(i \mid 0,1;0,1 \rangle + \mid 0,1;1,0 \rangle - \mid 1,0;0,1 \rangle + i \mid 1,0;1,0 \rangle\right).
\end{equation}  
Here, the state is given in the four mode photon number basis, where the first two modes form system A and the final two modes form system B. After the mode swap is applied, the state is
\begin{equation}
\mid \psi_2 \rangle = \frac{1}{2}\left(i \mid 0,0;1,1 \rangle + \mid 0,1;1,0 \rangle - \mid 1,0;0,1 \rangle + i \mid 1,1;0,0 \rangle\right).
\end{equation}  
As a result of the swap operation, system A is now entangled with system B, where each system exists in a four-dimensional subspace of its Hilbert space given by the states $\{\mid 0,0\rangle, \mid 0,1 \rangle, \mid 1,0 \rangle, \mid 1,1 \rangle\}$. However, the states $\mid 0,0 \rangle$ and $\mid 1,1 \rangle$ are not part of the qubit subspace $\{\mid 0,1 \rangle, \mid 1,0 \rangle\}$. It is therefore convenient to post-select only outcomes where one photon is in A and the other is in B. This post-selection can be represented by a projection operator $\hat{\Pi}$ and a post-selection probability $P_\Pi$, so that the normalized output of the post-selected swap operation is
\begin{equation}
\label{eq:swap}
\frac{1}{\sqrt{P_\Pi}}\;\hat{\Pi}\mid \psi_2 \rangle = \frac{1}{\sqrt{2}} \left(\mid 0,1;1,0 \rangle - \mid 1,0;0,1 \rangle \rangle\right).
\end{equation}  
This is a maximally entangled state of the two qubits A and B, where any linear optics transformation applied in parallel to A and to B will result in perfectly anti-correlated outcomes. 

\section{Non-local two-photon interferences}

Eq.(\ref{eq:swap}) represents the entangled state as a superposition of a Fock state with no photons in the swapped modes ($\mid 1,0;0,1 \rangle$) and a Fock state with both photons in the swapped modes ($\mid 0,1;1,0 \rangle$). These two photon number distributions interfere because the indistinguishability of photons means that we cannot know whether the photons were swapped or not. As a result, the outcomes of interferences between the swapped paths and the unswapped paths will be correlated as well, with photons appearing either in the two inner paths or the two outer paths of the four output paths shown in Fig. \ref{setup}. 

To better understand the physics of the effect, it may be useful to trace the propagation of the input modes $\hat{a}_1(\mathrm{in})$ and $\hat{b}_2(\mathrm{in})$ through the interferometer. Since the corresponding creation operators can be used to represent the input photons, it is convenient to express the input mode creation operators as linear superpositions of the output mode creation operators. The output mode terms can be arranged to reveal the origin of the nonlocal Hong-Ou-Mandel effect,
\begin{eqnarray}
\label{eq:grouping}
   \hat{a}^\dagger_1(\mathrm{in}) &=& - \frac{1}{2}\left(\hat{a}^\dagger_1(\mathrm{out}) - \hat{b}^\dagger_2(\mathrm{out})\right)  + \frac{i}{2} \left(\hat{a}^\dagger_2(\mathrm{out}) + \hat{b}^\dagger_1(\mathrm{out})\right),
   \nonumber \\
   \hat{b}^\dagger_2(\mathrm{in}) &=& \frac{1}{2}\left(\hat{a}^\dagger_1(\mathrm{out}) - \hat{b}^\dagger_2(\mathrm{out})\right) 
 + \frac{i}{2} \left(\hat{a}^\dagger_2(\mathrm{out}) + \hat{b}^\dagger_1(\mathrm{out})\right).
 \end{eqnarray}
Both input modes can be written as orthogonal superpositions of only two modes, where both of the modes are given by equal superpositions of one mode from A and one mode from B. Multi-photon interference ensures that both photons will be output in the same mode,
\begin{eqnarray}
\hat{a}^\dagger_1(\mathrm{in}) \hat{b}^\dagger_2(\mathrm{in}) \mid \mbox{vac.}\rangle &=&
- \frac{1}{4} \left(\hat{a}^\dagger_1(\mathrm{out}) - \hat{b}^\dagger_2(\mathrm{out})\right)^2 \mid \mbox{vac.}\rangle
\nonumber \\ &&
- \frac{1}{4} \left(\hat{a}^\dagger_2(\mathrm{out}) + \hat{b}^\dagger_1(\mathrm{out})\right)^2 \mid \mbox{vac.}\rangle.
\end{eqnarray}
Note that the two nonlocal modes are superpositions of different local modes. When one photon is detected in A and the other is detected in B, the detection of a photon in $\hat{a}_1$ will always coincide with a detection of a photon in $\hat{b}_2$, and the detection of a photon in $\hat{a}_2$ will always coincide with a detection of a photon in $\hat{b}_1$.
\begin{equation}
  \hat{\Pi}  \; \hat{a}^\dagger_1(\mathrm{in}) \hat{b}^\dagger_2(\mathrm{in}) \mid \mbox{vac.}\rangle = \frac{1}{2}\left(\hat{a}^\dagger_1(\mathrm{out}) \hat{b}^\dagger_2(\mathrm{out}) \mid \mbox{vac.}\rangle-  \hat{a}^\dagger_2(\mathrm{out}) \hat{b}^\dagger_1(\mathrm{out})\mid \mbox{vac.}\rangle\right)
\end{equation}
Mode swapping thus implements a nonlocal version of the conventional Hong-Ou-Mandel effect, where the post-selection of one output photon in each two mode system ensures that the photons are detected in spatially separated components of the same mode. Oppositely, the probabilities of detecting photons in $\hat{a}_1$ and $\hat{b}_1$ or in $\hat{a}_2$ and $\hat{b}_2$ are close to zero.

\section{Wave-particle dualism in multi-photon optics}

Multi-photon interferences are collective interferences of indistinguishable particles, making it very difficult (and perhaps somewhat pointless) to explain them in terms of individual particle paths. On the other hand, the representation of input photons by their creation operators provides a convenient compromise between the wave amplitudes represented by the operators and the observation of individual photons. As shown in Eq.(\ref{eq:grouping}), it is then possible to identify nonlocal interference effects by identifying the corresponding superposition of output modes in the creation operators of the input modes. Multi-photon interference effects can then be traced to their origin in the optical coherences described by the modes. 

It may be interesting to consider the possible applications of the multi-mode analysis described above to interferences of more than two photons. In particular, mode swapping operations can be used to generate and distribute entanglement between separate multi-mode systems, and non-local modes can be used to keep track of the non-classical correlations generated by such operations. Optical quantum information processes can then be described efficiently in terms of interacting multi-mode multi-photon systems, where information is encoded in the photon number distribution over the local set of modes. 

\section{Conclusions}
We have shown that mode swap operations between two multi-mode systems can result in non-local multi-photon interference effects when the local photon number is not changed by the mode swap operation. The best way to understand this quantum optical nonlocality is through the identification of nonlocal modes. In the non-local version of the Hong-Ou-Mandel effect, both photons exit the interferometer in the same non-local output mode described by an equal superposition of a mode in A with a mode in B. Even though the photons are detected in spatially separated locations, it is possible to identify this effect as a form of non-local photon bunching, where the separately detected photons can be traced back to the same original mode. 

The implementation of nonlocal multi-photon interference by a simple mode swapping operation is both important for the efficient construction of large scale quantum optical networks and for a better understanding of the physics of multi-photon multi-mode systems. The analysis presented above provides a promising starting point for a systematic exploration of the vast possibilities opened up by recent advances in the field of optical quantum computation and quantum information processing.

\acknowledgments % equivalent to \section*{ACKNOWLEDGMENTS}       
This work was supported by ERATO, Japan Science and Technology Agency (JPMJER2402). JRH acknowledges support from a Royal Society Research Grant (RG/R1/251590), and from their EPSRC Quantum Technologies Career Acceleration Fellowship (UKRI1217).

\end{document}